\documentclass[lettersize,journal]{IEEEtran}
\usepackage{amsmath,amsfonts}
\usepackage{algorithmic}
\usepackage{array}
\usepackage{textcomp}
\usepackage{stfloats}
\usepackage{url}
\usepackage{verbatim}
\usepackage{graphicx}
\usepackage{subcaption}
\def\BibTeX{{\rm B\kern-.05em{\sc i\kern-.025em b}\kern-.08em
    T\kern-.1667em\lower.7ex\hbox{E}\kern-.125emX}}
\usepackage{balance}
\usepackage{bm}
\usepackage{booktabs}
\usepackage{multirow}
\usepackage{graphicx}

\begin{document}
\title{ Autoencoder-assisted Feature Ensemble Net for Incipient Faults}
\author{Mingxuan Gao, Min Wang,~\IEEEmembership{Member,~IEEE}, Maoyin Chen,~\IEEEmembership{Member,~IEEE}
\thanks{This work was supported by the National Natural Science Foundation
	of China under Grant 61873143 (Corresponding author: Maoyin Chen).}

\thanks{Mingxuan Gao is with the Institue of education, Tsinghua University,
	Beijing 100084, China.}

\thanks{Min Wang is with the Department of Advanced Design and Systems Engineering, City University of Hong Kong, Hong Kong and the School of Automation Engineering, University of Electronic Science and Technology of China, Chengdu 611731, China.}

\thanks{Maoyin Chen is with the Department of Automation, Tsinghua University, Beijing 100084, China (e-mail: mychen@tsinghua.edu.cn).}
}

%\markboth{Journal of \LaTeX\ Class Files,~Vol.~18, No.~9, September~2020}%
%{How to Use the IEEEtran \LaTeX \ Templates}
\bibliographystyle{IEEEtran} 

\maketitle

\begin{abstract}
Deep learning has shown the great power in the field of fault detection. However, for incipient faults with tiny amplitude, the detection performance of the current deep learning networks (DLNs) is not satisfactory. Even if prior information about the faults is utilized, DLNs can’t successfully detect faults 3, 9 and 15 in Tennessee Eastman process (TEP). These faults are notoriously difficult to detect, lacking effective detection technologies in the field of fault detection. In this work, we propose Autoencoder-assisted Feature Ensemble Net (AE-FENet): a deep feature ensemble framework that uses the unsupervised autoencoder to conduct the feature transformation. Compared with the principle component analysis (PCA) technique adopted in the original Feature Ensemble Net (FENet), autoencoder can mine more exact features on incipient faults, which results in the better detection performance of AE-FENet. With same kinds of basic detectors, AE-FENet achieves a state-of-the-art average accuracy over 96\% on faults 3, 9 and 15 in TEP, which represents a significant enhancement in performance compared to other methods. Plenty of experiments have been done to extend our framework, proving that DLNs can be utilized efficiently within this architecture.
\end{abstract}


\section{Introduction}
\IEEEPARstart{W}{ith} the advancement of information technology, modern industrial systems have become larger in scale and more complex in structure \cite{ge2017review}, which greatly increases the probability of system failure. During the operation of the system, incipient faults may result in irremediable consequences.In practical industrial systems, significant accidents frequently emerge as a result of incipient faults. In order to avoid huge economic losses and heavy casualties \cite{amar2014vibration, demetriou1998incipient, wangmin2022TSMCAAnomaly}, incipient faults should be detected as soon as possible.

Detecting incipient faults can be challenging, particularly when they are affected by unknown disturbances or masked by other factors \cite{ren2010single}. The presence of widespread feedback controls also poses a significant challenge in detecting the aforementioned nascent faults, compounded by their minuscule magnitudes \cite{wangmin2022AutomaticaHybrid}. Some particular fault types, including faults 3, 9 and 15 in Tennessee-Eastman process (TEP), have gained extensive recognition as prototypical incipient faults, known for their notably difficult detectability \cite{cao2021hierarchical}, \cite{ding2009subspace}.

Over the past few decades, a range of methods have been proposed to address faults 3, 9, and 15 in TEP. These methods encompass various approaches, mainly including multivariate statistical analysis \cite{ku1995disturbance}, \cite{shams2011fault}, machine learning \cite{he2007fault}, \cite{zhang2018fault} and deep learning \cite{zhang2021sparsity}. Based on multivariate statistical analysis, Mustapha\textit{ et al.} proposed a modified moving window PCA with fuzzy logic filter \cite{ammiche2018combined}, which provides an fixed PCA model using adaptive thresholds. Wahiba\textit{ et al.} proposed a distribution dissimilarity technique named Kullback-Leibler (KL) divergence through principle component analysis (PCA) \cite{bounoua2020online}. 
Unfortunately, few multivariate statistical analysis methods can detect tiny faults successfully.

With the aim to develop a kind of unsupervised DLN, the recent progress on DLNs is discussed \cite{wangmin2022TIIRecursive}. Wang\textit{ et al.} proposed a deep learning model generated by variational autoencoder probability \cite{wang2021supervised}. Sun\textit{ et al.} introduced a Bayesian recurrent neural network with variational difference \cite{sun2020fault}, which can handle complex nonlinear dynamics. Additionally, Zhao\textit{ et al.} presented global and local structure-based neural networks \cite{zhao2019global}, which can adaptively train neural networks while considering both global variance information and local geometric structures. Despite employing the aforementioned methods, the performance of fault detection on faults 3, 9, and 15 in the TEP dataset has fallen profoundly below expectations.

In contrast to unsupervised deep learning networks (DLNs), supervised DLNs exhibit superior performance. Zhang\textit{ et al.} proposed a scalable deep belief network-based model \cite{zhang2017deep}, which utilizes sub-networks to extract the features of faults in both time and space dimensions. Using quantum computing to assist in the generative training process, Ajagekar\textit{ et al.} displayed a fault diagnosis model based on quantum computing \cite{ajagekar2020quantum}. Wu\textit{ et al.} utilized deep convolutional neural network for fault diagnosis \cite{wu2018deep}. A temporal architecture is adopted in the work of Lomov\textit{ et al.}, while augmenting the training data with generative adversarial networks \cite{lomov2021fault}, resulting in an overall performance boost. If temporal deep learning is employed, the detection rates on faults 3, 9, 15 in TEP can reach 96\%, 92\%, and 70\%. In summary, the detection performance by unsupervised deep learning still remains poor, which is incongruous with the substantial capabilities of deep learning.

In our recent work, a novel deep framework called the Feature Ensemble Network (FENet) was developed by our research group. Its primary objective is to efficiently detect faults at positions 3, 9, and 15 in TEP dataset using an unsupervised learning approach, as described in \cite{DechengLiu2022}. To construct the input feature layer, we integrate the features extracted by basic detectors. Subsequently, the feature matrix obtained from the previous stage undergoes a transformation process in the hidden feature transformer layers, combining sliding-window patches and PCA. In the final step, all feature matrices in the last hidden layer are stacked together to create a comprehensive feature matrix. This matrix allows for the design of a detection index using normalized singular values derived from sliding windows.

\begin{figure*}
	\begin{center}
		\setlength{\abovecaptionskip}{0pt}
		\setlength{\belowcaptionskip}{0pt}
		\centering
		\includegraphics[width=1.99\columnwidth]{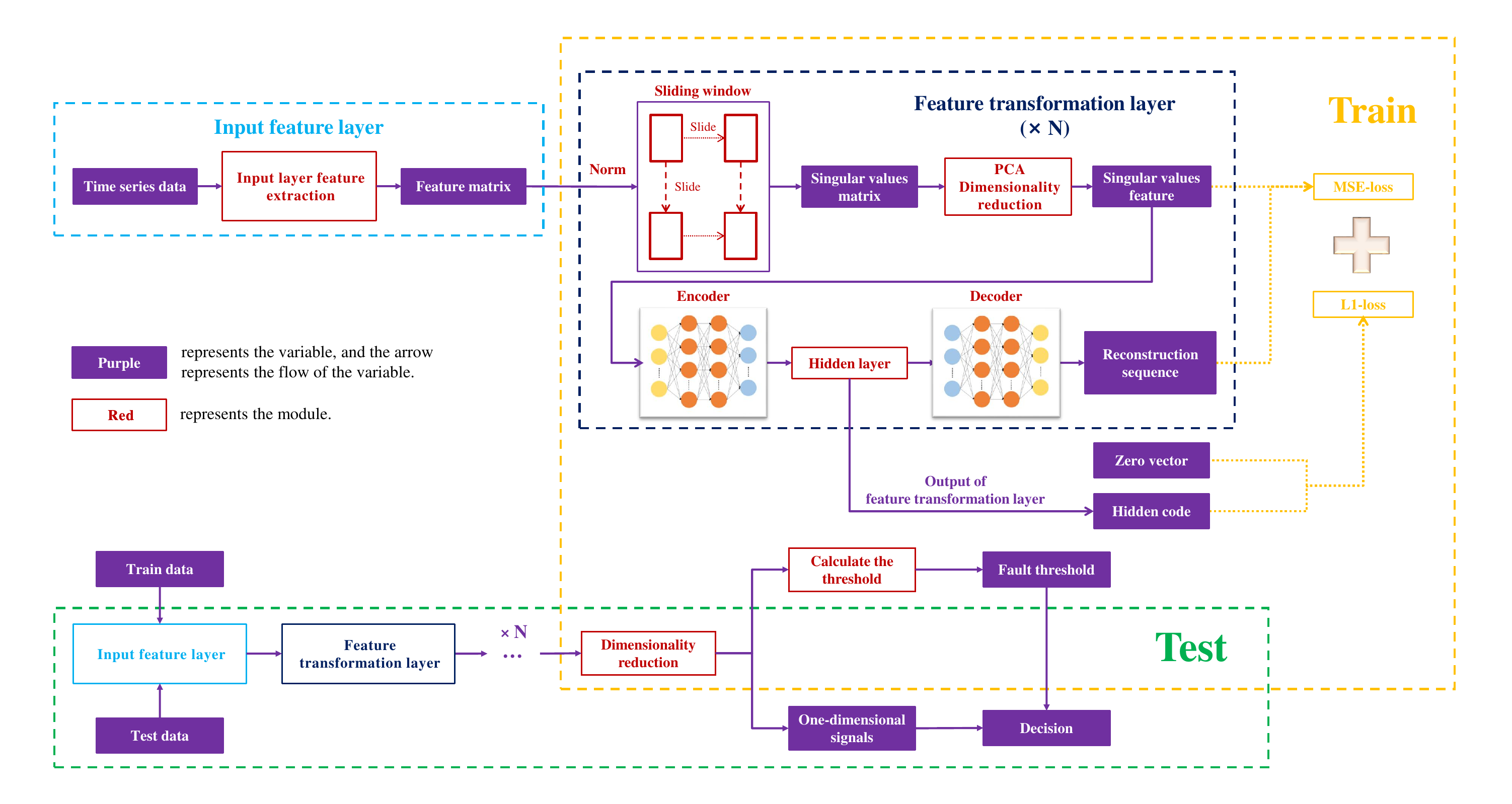}
	\end{center}
	\caption{The general framework of AE-FENet}
	\label{AE-FENet architecture}
\end{figure*}

Despite being categorized as a deep framework, FENet does not fully qualify as a DLN due to its reliance solely on PCA as the feature miner for detecting early-stage faults, without incorporating neural networks. Therefore, the fault feature extraction accuracy of FENet is compromised, restricting the potential for further enhancements in its detection performance. In this paper, we propose an unsupervised DLN named Autoencoder-assisted FENet (AE-FENet), wherein we substitute PCA with an autoencoder within the FENet framework. The primary contributions of AE-FENet can be summarized as follows:

\begin{enumerate}
\item[1)] \textbf{A deep feature ensemble framework with unsupervised DLNs.} The proposed framework, AE-FENet, employs an autoencoder-based feature transformation approach to construct a DLN, which surpasses the conventional deep framework. It greatly improves the performance of unsupervised DLNs for incipient faults, bridging the existing methodological gap.

\item[2)] \textbf{The superior detection performance to FENet.} By leveraging the autoencoder in AE-FENet, which captures more precise characteristics of incipient faults when contrasted with PCA employed in FENet, the former achieves enhanced average performance in detecting faults 3, 9, and 15 within the TEP dataset, utilizing the same set of fundamental detectors in both frameworks.

\item[3)] \textbf{The high scalability of feature transformers.} The autoencoder, serving as a feature transformer, can be extended to encompass various types of autoencoders, including sparse autoencoders, variational autoencoders, and attention-based autoencoders. Regardless of the specific type of autoencoder employed, the overall performance in detecting incipient faults remains excellent, which validates the practical value of DLNs within the AE-FENet framework for detecting incipient faults.

\end{enumerate}

The remainder of the paper is structured as follows. Section \uppercase\expandafter{\romannumeral2} introduces the problem of incipient fault detection. The structure of AE-FENet is illustrated in Section \uppercase\expandafter{\romannumeral3}. In Section \uppercase\expandafter{\romannumeral4}, simulation of TEP are carried out, and the scalability of AE-FENet are proposed, displaying the performance superiority of DLNs. The conclusions are presented in Section \uppercase\expandafter{\romannumeral5}.

\section{Problem formulation}
\label{Section2}
Although the detection of incipient faults has made great progress, the successful detection of particularly challenging faults, specifically faults 3, 9, and 15 in TEP, remains scarce. Even state-of-the-art DLNs have not met the expected detection performance for these incipient faults, whether in supervised or unsupervised contexts.

In this research paper, our objective is to develop an unsupervised DLN capable of efficiently detecting incipient faults, including faults 3, 9, and 15 in TEP. Our proposed DLN will be built upon a recently introduced deep framework called FENet, which, strictly speaking, is not categorized as a DLN. We will incorporate neural networks such as autoencoders into the original FENet architecture to facilitate feature transformation and fault feature extraction. This integration is expected to improve the detection performance of incipient faults when compared to the original FENet framework.

\section{Autoencoder-assisted FENet(AE-FENet)}
\label{Section3}
By implementing a multi-layer deep network structure, AE-FENet demonstrates an enhanced capability in detecting incipient faults when compared to FENet. It is important to note that FENet comprises several layers, namely the input feature layer, feature transformation layers, output feature layer, and decision layer, as described by Liu et al. in their publication \cite{DechengLiu2022}. Specifically, within FENet, the extraction of deep features related to incipient faults is accomplished through hidden feature transformers employing sliding windows and PCA techniques.
However, PCA exhibits relatively limited efficacy in extracting fault features, thereby constraining the potential for further improvements in the detection performance of FENet. To address this limitation, we introduce the renowned autoencoder as an alternative feature transformer for extracting incipient fault features. This involves replacing PCA with an autoencoder in the hidden feature transformation layers of AE-FENet. The overall architecture of AE-FENet is illustrated in Fig.\ref{AE-FENet architecture}.

\subsection{The input layer}
Various fault detection techniques enable the extraction of diverse features from the data , including Gaussian features, sliding window features, and others \cite{DechengLiu2022}. Within the input feature layer, the integration of different types of fundamental detectors enhances the efficiency of data feature extraction.

Let $ \bm{X}=[\bm{x}_{1},\bm{x}_{2},..., \bm{x}_{n} ]^{T} $ $\in$ $ \mathbb{R}^{m} $ be the training set with only normal data, where $n$ ($n>m$) is the number of samples. The outputs of the basic detectors are denoted as $f(x)$, and the detection feature vectors are constructed. They can be stacked into a feature integration matrix as follows \cite{DechengLiu2022}:
\begin{equation}
\bm{U} = \begin{bmatrix}u_{1} & u_{2} & ... & u_{n}\end{bmatrix}^{T} \in \mathbb{R} ^{n \times k} 
\end{equation}
where $
u_{i}=[f_{1}(\bm{x}_{i}),f_{2}(\bm{x}_{i}),...,f_{k}(\bm{x}_{i})]^{T} \in \mathbb{R}^{k}   
$.

\subsection{The feature transformation layer}
\begin{figure*}
	\begin{center}
		\setlength{\abovecaptionskip}{0pt}
		\setlength{\belowcaptionskip}{0pt}
		\centering
		\includegraphics[width=1.99\columnwidth]{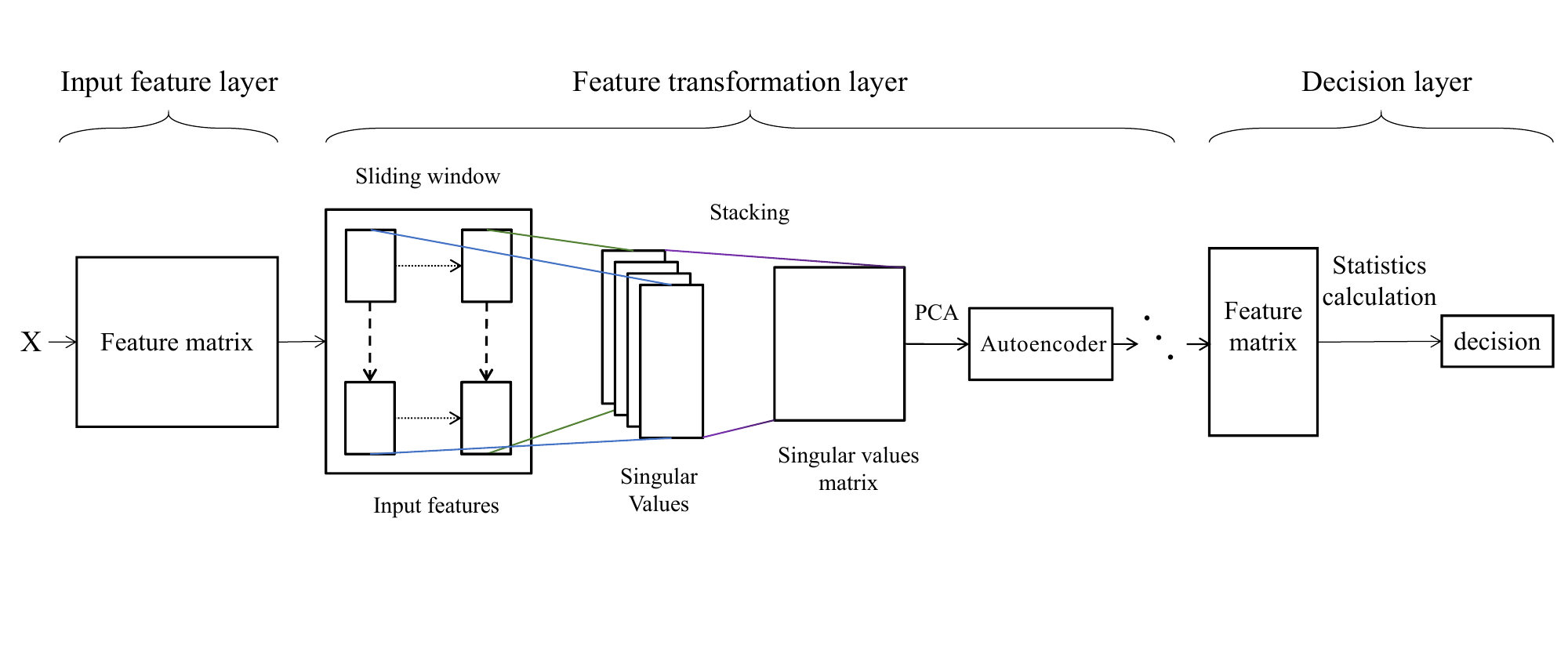}
	\end{center}
	\caption{The network structure of AE-FENet}
	\label{structure}
\end{figure*}

The feature transformation layer comprises three sequential steps, namely: computation of sliding window employing singular values, dimensionality reduction using PCA, and  application of an autoencoder. Fig.\ref{structure} illustrates the network structure of AE-FENet.

Feature ensemble matrix is expressed as $\bm{U}$, and the width of sliding window is set to $\omega$ ($\omega>k$). Through a single-step sliding window, the matrix is obtained as follows:
\begin{equation}
\bm{U}_{q} 
=\left[\bm{u}_{q-\omega+1},\bm{u}_{q-\omega+2},...\bm{u}_{q} \right]^{T} \in \mathbb{R}^{n \times k}
\end{equation}
where $q - \omega + 1$, $q - \omega + 2$, ..., $q$ indicate the sampling instants of the process data orderly. The column vector is as follows:
\begin{equation}
\bm{u}_{q-\omega+1} = \left[f_{1}(\bm{x}_{q-\omega+1}),
f_{2}(\bm{x}_{q-\omega+1}),...f_{k}(\bm{x}_{q-\omega+1})\right].
\end{equation}

The number of feature transformation layers is assumed to $l^{max}$ in the framework. When $l^{max} \ge 1$, $l=l^{max}-1$. After reducing the dimension with sliding window, the matrix in the $l+1$ feature transformation layer is expressed as $\bm{U^{l+1}}$.

There are $m_{l}$ detection indicators and $n_{l}$ sampling moments in the feature integration matrix. Mutiple sliding window matrices of different sizes are contained in each feature transformation layer. For simplicity, the widths of sliding windows are all set to $\omega$, then $n_{l+1}=n_{l}-\omega+1$. The length of the sliding windows is set to $h_{l}$, representing that $h_{l}$ basic detectors are randomly selected from $m_{l}$ detection indicators, where $0<h_{l}$ $\le$ $ m_{l}$.

Here $u_{i,j}^{l}$ is used to represent the $i$-th row and $j$-th column of the feature matrix $\bm{U^{l}}$, and $h_{l}$ column elements are selected from the $m_{l}$ columns randomly. Then, the sliding window matrix corresponding to the sample $\bm{x_{q}}$ is expressed as \cite{DechengLiu2022}:
\begin{equation}
\bm{U^{l}_{q}} = \begin{bmatrix} u^{l}_{q-\omega+1,1} & ... & u^{l}_{q-\omega+1,h_{l}}\\
\vdots & \cdots & \vdots \\
u^{l}_{q,1} & ... & u^{l}_{q,h_{l}} \end{bmatrix}
\in \mathbb{R}^{\omega \times h_{l}}
\end{equation}
where $q = n-n_{l}+\omega,n-n_{l}+\omega+1,...,n $.

The technique of normalization of the overall mean and variance is adopted. The matrix after normalization is expressed as $\overline{\bm{U}}^{l}_{q}$. According to the different sampling time, $\overline{\bm{U}}^{l}_{q}$ is dealt with singular value decomposition, and the singular value $\bm{\sigma_{q}}^{l}$ is calculated ($\bm{\sigma_{q}}^{l} \in \mathbb{R}^{h_{l}}$). The singular value matrix corresponding to different samples is obtained \cite{DechengLiu2022}:
\begin{equation}
\bm{V}^{l}_{c} = \left[\bm{\sigma}^{l}_{n-n_{l}+\omega},\bm{\sigma}^{l}_{n-n_{l}+\omega+1},...,\bm{\sigma}^{l}_{n} \right]^{T}.
\end{equation}	

To obtain the fusion singular value matrix $\bm{V}^{l}$, for $c=1, 2, ..., C^{h_{l}}_{m_{l}}$, the singular value matrix $\bm{V}^{l}_{c}$ is stacked by:
\begin{equation}
\bm{V}^{l} = \left[\bm{V}^{l}_{1},\bm{V}^{l}_{2},..., \bm{V}^{l}_{C^{h_{l}}_{m_{l}}}\right]
\in \mathbb{R}^{(n_{l}-\omega+1) \times h_{l}}.
\end{equation}	

Note that the singular value matrix contains a certain amount of zero-value information. In order to reduce information redundancy, PCA is used to reduce the dimension of the singular value matrix \cite{ChenH2018}, \cite{ChanTH2015}.

Here $\bm{V}^{l}$ is the input for the process of PCA. With calculating the sample mean, the data can be centralized and the covariance matrix $\sum$ is calculated. After performing eigenvalue decomposition on $\sum$, the eigenvectors corresponding to the largest $t$ eigenvalues are used to form a projection matrix. With the projection matrix to project matrix $\bm{V}^{l}$, in the feature transformation layer, the singular value matrix $\bm{V}^{l'}$ is obtained, which is the input data layer of autoencoder.

The choice of encoder function $f_{\vartheta}$ and decoder function $g_{\vartheta}$ depends on the scope and nature of the input domian. In this paper, affine functions of encoder and decoder are showed as follows:
\begin{equation}
f_{\vartheta_f}(x) = sigm(W\bm{x}+b)
\end{equation}	
\begin{equation}
g_{\vartheta_g}(h) = sigm(W^{'}\bm{h}+d)
\end{equation}
where $sigm$ is the sigmoid activation function, $W$ and $W^{'}$ are weights, $b$ and $d$ are bias. The set of parameters is represented as $\vartheta=\{W, b, W^{'}, d\}$. Using $w_{0}$ to represent the initial value, $w$ in each dimension can be updated by:
\begin{equation}
w_{i+1} = w_{i}-\alpha \frac{\partial f(w)}{\partial w}.
\end{equation}	

The autoencoder employs two loss functions, namely mean squared error (MSE) and mean absolute error (MAE), also referred to as MSE-loss and L1-loss respectively. MSE-loss serves the purpose of enforcing input samples to conform closely to the reconstructed sequences, thereby facilitating the autoencoder's ability to achieve accurate reconstructions of the input sequences. On the other hand, L1-loss plays a critical role in constraining the hidden layer coding to zero, which ensures that the distribution of hidden information in normal data remains restricted.

For data $D(x, y)$ containing \emph{N} samples, $x$ represents the output distribution of the neural network, and $y$ stands for the distribution sequence of real features. The dimension of $x$ is same as that of $y$, then the $i$-th sample loss calculation expressions are as follows:
\begin{equation}
\mathcal{L}^{MSE}_{i} = \frac{1}{N}(x_{i}-y_{i})^{2},
\end{equation}	
\begin{equation}
\mathcal{L}^{L1}_{i} = \frac{1}{N}|h_{i} - 0|.
\end{equation}

In practical tasks, autoencoder is often composed of multi-layer perceptrons. $F(\cdot,\varTheta_F)$ represents the encoders $f_{\vartheta_f^1}$, $f_{\vartheta_f^2}$, ..., $f_{\vartheta_f^m}$, which is the composite function of the $m$ encoding functions. And $G(\cdot, \varTheta_G)$ represents the decoders  $g_{\vartheta_g^1}$, $g_{\vartheta_g^2}$, ..., $g_{\vartheta_g^n}$, which is the composite function of the $n$ decoding functions.

Note that the autoencoder is expressed as follows:
\begin{equation}
y = G(F( x , \varTheta_F), \varTheta_G).
\end{equation}

Using $\varTheta = \{\varTheta_F, \varTheta_G\}$ to represent the overall parameters of the network, the optimization problem constrained by the loss functions $\mathcal{L}^{MSE}$ and $\mathcal{L}^{L1}$ is illustrated by:
\begin{equation}
\begin{aligned}
\varTheta^* = \arg\min_{\varTheta} 
\mathbb{E}_x &
\mathcal{L}^{MSE} ( G(F(x, \varTheta_F), \varTheta_G), x) \\
&+ \mathcal{L}^{L1} ( F(x, \varTheta_F), 0 ) 
\end{aligned} .
\end{equation}

\begin{figure*}
	\centering
	\includegraphics[width=14cm]{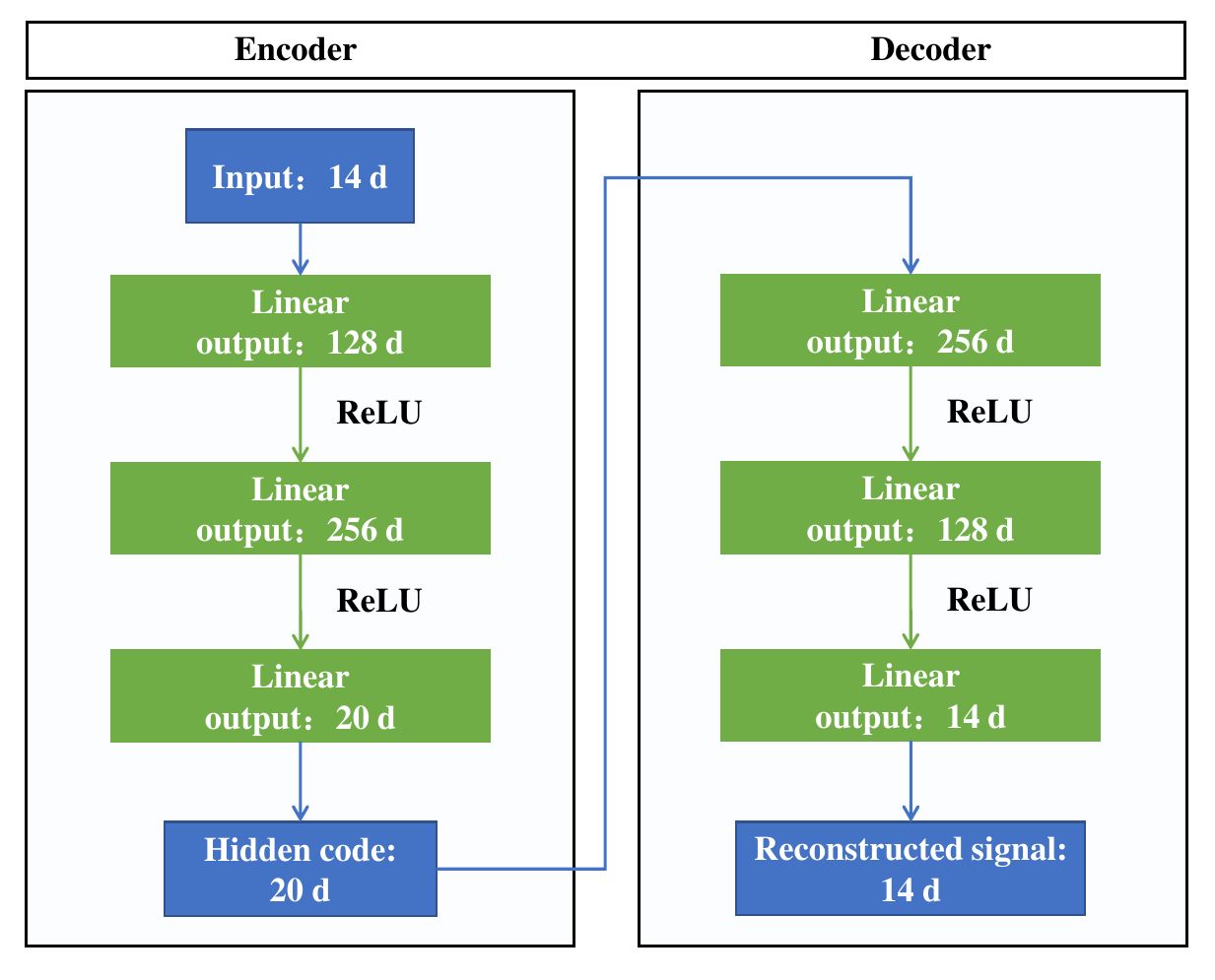}
	\caption{The structure of autoencoder}
	\label{Linear}
\end{figure*}

The AE-FENet model employs a linear transformation for configuring the fully connected layer, while ReLU serves as the activation function within the network. The schematic representation of this component of the model architecture can be observed in Fig.\ref{Linear}, referred to as "Linear" within the figure. Here, the variable $d$ succinctly denotes the dimension. The model incorporates three linear network layers, with the input being the matrix resulting from PCA dimension reduction. Following the encoding and decoding processes, the ultimate output dimension is constrained to 20.

By setting up multiple feature transformation layers, the unsupervised DLN (AE-FENet) is constructed. As the network advances through the deeper layers, each layer progressively captures abstract representations in a more comprehensive manner.

\subsection{The decision layer}
The decision layer takes as input the output feature matrix $\bm{U}^{0}$, which is generated by the preceding feature transformation layer. The hidden layer produced by the final encoder layer is donated as $\sigma^{0}_{q}$. In this context, $\mu$ and $std$ represent the mean and the standard deviation of normal samples respectively. Besides, $p$ represents the $p$-norm of the vector. Then the detection index of the sample $\sigma^{0}_{q}$ can be formulated as follows:

\begin{equation}
D_{q} = \left\|\frac{ \sigma^{0}_{q} - \mu}{std} \right\|_{p}.
\end{equation}	

The calculation of $D^{lim}$, the control limit of the detection index, is determined based on the provided confidence level. AE-FENet employs L1-loss during the training of the autoencoder structure to regulate the distribution of the encoder's hidden layer output. Consequently, when utilizing the hidden features from the final layer to compute the statistics, the approach of utilizing the one-norm ($p=1$) is implemented.

\section{Simulation}
\label{Section4}

\subsection{Tennessee Eastman process} 
Tennessee Eastman process (TEP) is an open benchmark platform that is widely used to verify the performance of different detection methods \cite{zhang2020incipient}. Originally developed by the Eastman Chemical Company of the United States in 1993 \cite{downs1993plant}, this simulation platform generates data that accurately represents the time-varying, strongly coupled, and nonlinear characteristics of real chemical reaction processes. TEP encompasses 53 process monitoring variables, including 22 continuous variables, 19 process variables, and 12 manipulated variables \cite{DechengLiu2022}. The dataset provided for TEP consists of both a training set and a test set, encompassing a total of 21 types of predetermined disturbances.

For the simulation experiments, the training set is composed solely of normal data, while the test set comprises test samples. The training phase and the test phase each ran for a duration of 200 hours, utilizing data generated from the simulation model \cite{bathelt2015revision}. The sampling interval for the samples is set at 3 minutes. In each test dataset, a fault is introduced after the 100th hour of simulation. The training data exclusively consists of normal data and has a sequence length of 4000. Similarly, the test data sequence also has a length of 4000, with the first 2000 entries representing normal data and the subsequent 2000 entries representing fault data. It is important to note that, due to the addition of fault 6 information, the simulation system can only operate for 7 hours \cite{WuH2018}, resulting in a distinct sequence length compared to the other datasets.

\subsection{Experiment setting and results}
The base detectors employed in this study comprise PCA, Dynamic Principal Component Analysis (DPCA), and three variants of the Mahalanobis Distance (MD), denoted as MD$^{1}$, MD$^{2}$, MD$^{3}$. These detector choices align with the fundamental detectors utilized in the FENet framework \cite{DechengLiu2022}. Consequently, the input feature layer is designed to capture a total of seven distinct features. Stated differently, the dimensionality of the feature integration matrix is seven.

In the AE-FENet architecture, the number of feature transformation layers is configured as two, and the widths of the sliding windows are both set to 150. Each autoencoder undergoes training for 10000 epochs on a single GPU, employing the Adam optimizer with a learning rate of 0.001. The decision layer incorporates a confidence level of 0.99.

Table \ref{PCA,DPCA,MD,KPCA,SparsePCA methods} lists the comparison of detection rates (FDRs) of PCA \cite{JoePCA}, DPCA \cite{ku1995disturbance}, MD \cite{mardia1977mahalanobis}, KPCA \cite{lee2004nonlinear} and Sparse PCA \cite{grbovic2012decentralized}. Among them, Kernel PCA covers three kinds of kernel functions, namely poly, rbf, and cosine. From Table \ref{PCA,DPCA,MD,KPCA,SparsePCA methods}, the detection rates of faults 3, 9, 15, 16 and 21 are very low, meaning that none of those methods can detect these faults effectively.

\begin{table*}[!htbp]
	\centering
	\caption{FDRs(\%) of PCA, DPCA, MD, KPCA and Sparse PCA.}
	\begin{tabular}{ccccccccccc}
		\toprule[1.5pt]
		\multirow{2}*{Falut} & \multirow{2}*{Type} & \multirow{2}*{PCA} & \multirow{2}*{DPCA} & \multicolumn{3}{c}{MD} & \multicolumn{3}{c}{KPCA} & \multirow{2}*{Sparse PCA} \\
		\cmidrule(r){5-7} \cmidrule(r){8-10}
		
		~ & ~ & ~ & ~ & MD$^{1}$ & MD$^{2}$ & MD$^{3}$ & poly & rbf & cosine & ~ \\
		\midrule
		0 & Normal & 1.70\% & 2.10\% & 1.10\% & 0.70\% & 0.70\% & 0.60\% & 1.00\% & 0.45\% & 0.55\%\\ 
		
		1 & Step & 99.95\% & 99.95\% & 99.95\% & 99.90\% & 99.90\% & 1.45\% & 0.00\% & 0.00\% & 1.30\%\\ 
		
		2 & Step & 99.90\% & 99.80\% & 99.85\% & 99.65\% & 99.65\% & 99.45\% & 0.00\% & 0.00\% & 93.55\% \\ 
		
		3 & Step & 5.70\% & 10.40\% & 2.65\% & 2.60\% & 1.05\% & 0.00\% & 0.65\% & 0.05\% & 0.00\%\\ 
		
		4 & Step & 99.95\% & 99.95\% & 99.95\% & 2.00\% & 99.95\% & 0.00\% & 0.00\% & 0.00\% & 0.00\%\\ 
		
		5 & Step & 3.35\% & 4.00\% & 2.10\% & 1.35\% & 2.00\% & 0.00\% & 0.95\% & 0.00\% & 0.00\%\\ 
		
		6 & Step &100.00\% & 100.00\% & 100.00\% & 100.00\% & 100.00\% & 70.63\% & 0.00\% & 0.00\% & 65.73\%\\
		
		7 & Step &100.00\% & 100.00\% & 100.00\% & 4.65\% & 100.00\% & 100.00\% & 0.00\% & 0.00\% & 99.95\%\\ 
		
		8 & Random variation & 99.60\% & 99.65\% & 99.65\% & 99.65\% & 99.60\% & 41.50\% & 0.00\% & 0.00\% & 37.00\% \\ 
		
		9 & Random variation & 7.70\% & 12.85\% & 5.55\% & 3.75\% & 1.65\% & 0.05\% & 0.45\% & 0.05\% & 0.05\% \\ 
		
		10 & Random variation & 92.35\% & 94.75\% & 95.55\% & 91.55\% & 76.80\% & 12.30\% & 0.00\% & 0.00\% & 6.10\% \\ 
		
		11 & Random variation &98.20\% & 99.60\% & 98.95\% & 90.25\% & 94.45\% & 25.90\% & 0.00\% & 0.00\% & 17.00\%\\ 
		
		12 & Random variation & 46.50\% & 61.20\% & 51.50\% & 46.30\% & 22.70\% & 1.30\% & 0.00\% & 0.00\% & 1.05\%\\ 
		
		13 & Slow drift & 97.50\% & 97.35\% & 97.45\% & 97.55\% & 97.45\% & 34.45\% & 0.00\% & 0.00\% & 30.20\%\\ 
		
		14 & Sticking & 99.90\% & 99.90\% & 99.90\% & 99.90\% & 87.10\% & 8.95\% & 0.00\% & 0.00\% & 3.00\% \\ 
		
		15 & Sticking & 3.05\% & 2.25\% & 1.30\% & 0.95\% & 0.80\% & 0.05\% & 0.60\% & 0.05\% & 0.10\% \\ 
		
		16 & Unknowing & 1.80\% & 2.40\% & 0.55\% & 0.65\% & 0.65\% & 0.00\% & 0.80\% & 0.15\% & 0.00\%\\ 
		
		17 & Unknowing & 99.10\% & 99.15\% & 99.15\% & 99.15\% & 88.40\% & 9.30\% & 0.00\% & 0.00\% & 7.15\% \\ 
		
		18 & Unknowing & 84.95\% & 93.20\% & 87.10\% & 83.35\% & 14.90\% & 1.25\% & 0.00\% & 0.00\% & 0.30\% \\ 
		
		19 & Unknowing & 99.90\% & 99.85\% & 99.90\% & 58.90\% & 99.90\% & 43.25\% & 0.00\% & 0.00\% & 19.95\%\\ 
		
		20 & Unknowing & 99.30\% & 99.30\% & 99.40\% & 99.45\% & 98.70\% & 0.45\% & 0.00\% & 0.00\% & 0.20\%\\ 
		
		21 & Constant position & 2.90\% & 3.65\% & 1.70\% & 1.45\% & 1.65\% & 0.00\% & 1.15\% & 0.05\% & 0.00\%\\ 	
	%	Average & ~ & \\	
		\bottomrule[1.5pt]
	\end{tabular}
	\label{PCA,DPCA,MD,KPCA,SparsePCA methods}
\end{table*}

In this simulation, faults 3, 5, 9, 15, 16, and 21 in TEP are used to compare the performance. The FDRs for these TEP faults are provided in Table \ref{FENet}, with reference to FENet \cite{DechengLiu2022} and AE-FENet.

Upon examining faults 5, 15, 16, and 21, the FDRs of FENet are all below 90\%, and the average accuracy is 85.73\%. By comparison, when $l^{max}=2$, the average FDR of AE-FENet has reached 98.66\%, which shows the superiority of AE-FENet. For typical incipient faults 3, 9, 15 in TEP, the FDRs of AE-FENet reach 97.55\%, 97.55\%, 96.05\%. At the same time, the FDRs for faults 16 and 21 also exceed 97\%. It proves that AE-FENet has excellent performance, compared to the original FENet.

\begin{figure*}
	\centering
	
	\subcaptionbox{PCA($T^{2}$)}{\includegraphics[width=5.8cm]{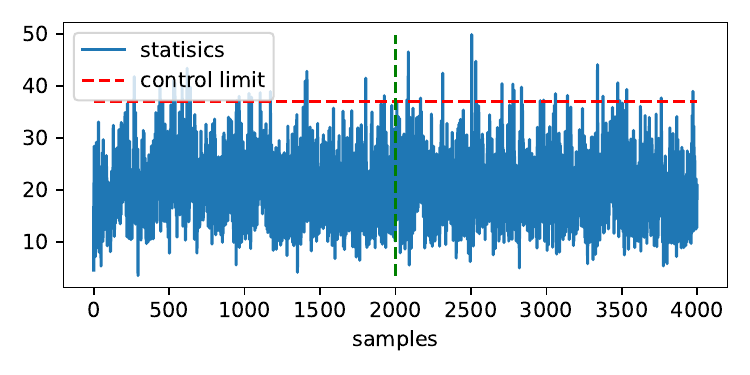}}
	\subcaptionbox{PCA($Q$)}{\includegraphics[width=5.8cm]{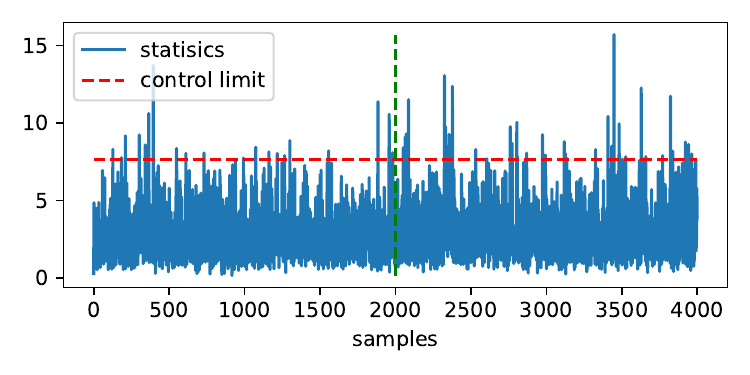}}
	\subcaptionbox{DPCA($T^{2}$)}{\includegraphics[width=5.8cm]{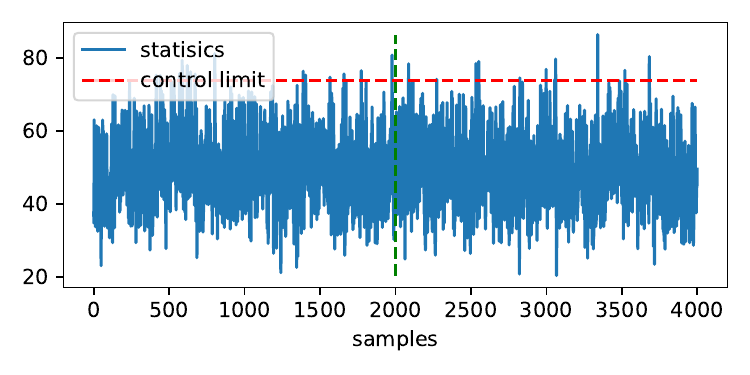}}
	
	\subcaptionbox{DPCA($Q$)}{\includegraphics[width=5.8cm]{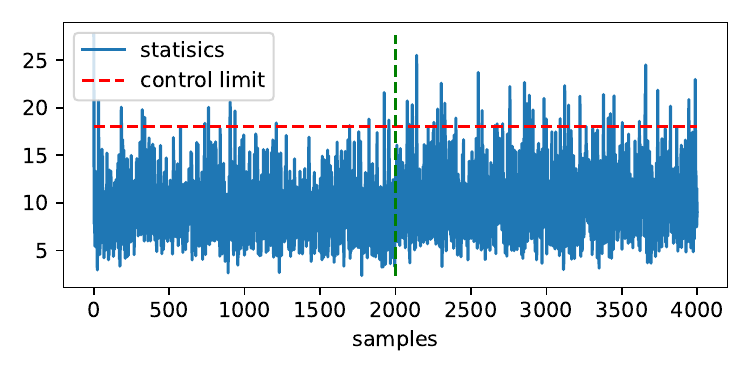}}
	\subcaptionbox{MD$^{1}$($d$)}{\includegraphics[width=5.8cm]{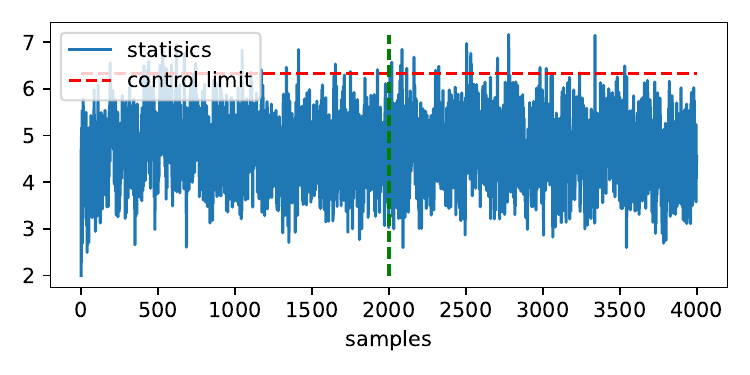}}
	\subcaptionbox{MD$^{2}$($d$)}{\includegraphics[width=5.8cm]{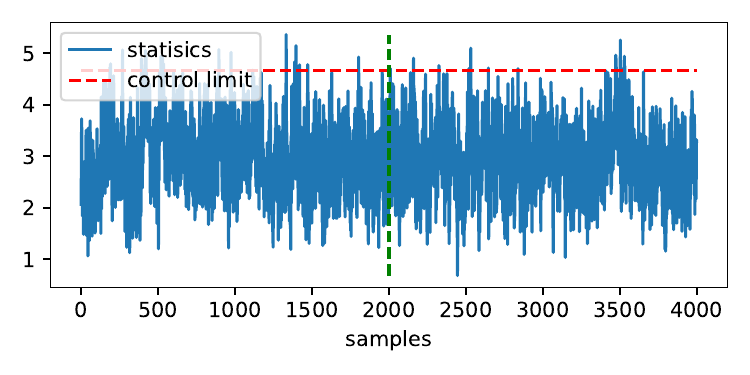}}
	
	\subcaptionbox{KPCA($poly$)}{\includegraphics[width=5.8cm]{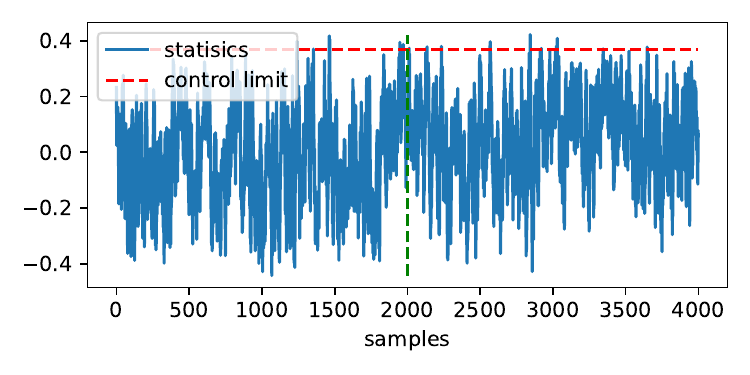}}
	\subcaptionbox{KPCA($rbf$)}{\includegraphics[width=5.8cm]{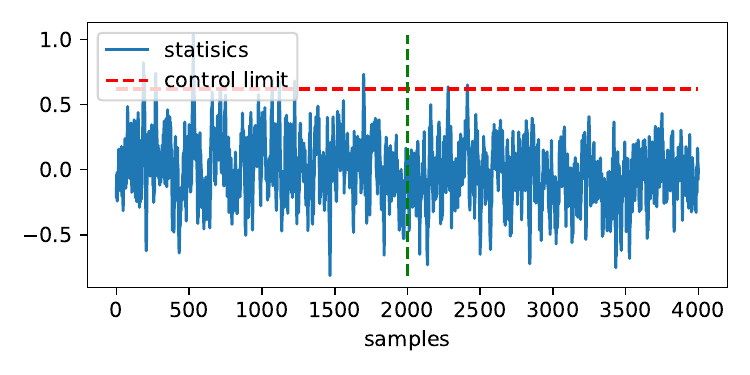}}
	\subcaptionbox{Sparse PCA}{\includegraphics[width=5.8cm]{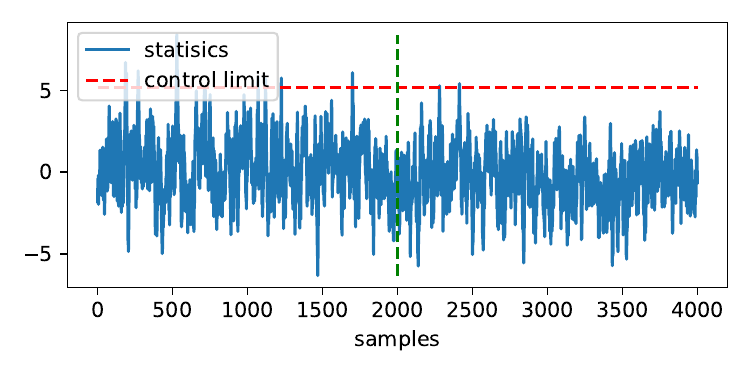}}
	
	\subcaptionbox{Layer 0}{\includegraphics[width=5.8cm]{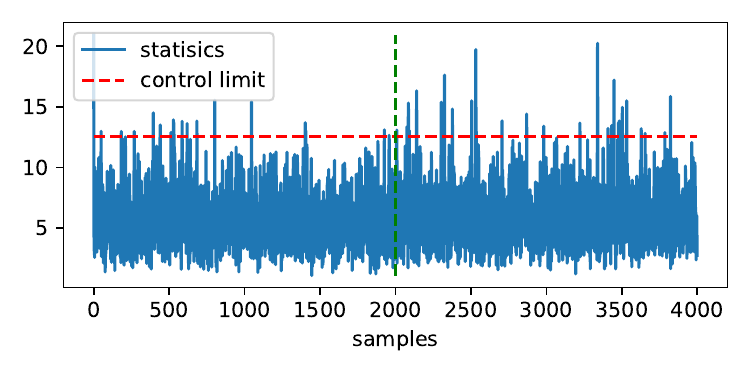}}
	\subcaptionbox{Layer 1}{\includegraphics[width=5.8cm]{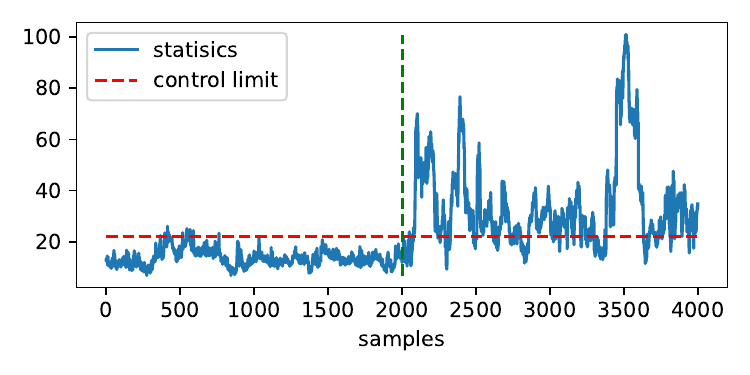}}
	\subcaptionbox{Layer 2}{\includegraphics[width=5.8cm]{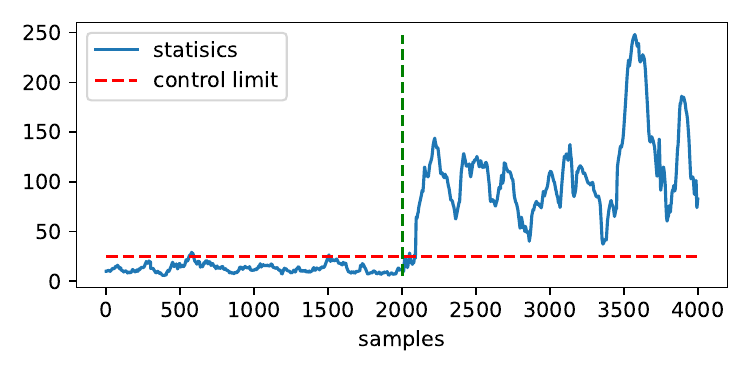}}
	\caption{Detection performance of basic detectors and AE-FENet for fault 15 in TEP}
	\label{ERR15}
\end{figure*}

Figure \ref{ERR15} presents a performance comparison between basic detectors and AE-FENet for TEP fault 15, with a fixed value of $\omega=150$. It can be observed that as the number of feature transformation layers increases, the performance of AE-FENet improves. This result provides intuitive evidence for the effectiveness of the feature transformation layer intuitively.

\begin{table*}
	\centering
	\caption{FDRs(\%) of FENet and AE-FENet}
	\begin{tabular}{ccccccc}
		\toprule[1.5pt]
		\multirow{2}*{Fault} & \multicolumn{3}{c}{FENet} & \multicolumn{3}{c}{AE-FENet} \\
		\cline{2-7}
		~ & $l^{max}$=0 & $l^{max}$=1 & $l^{max}$=2 & $l^{max}$=0 & $l^{max}$=1 & $l^{max}$=2 \\
		\midrule
		0 & 1.20\% & 1.40\% & 1.10\%  & 1.05\% & 0.20\% & 0.95\%\\
		1 & 99.90\% & 99.85\% & 99.85\% & 99.95\% & 99.90\% & 99.80\%\\
		2 & 99.65\% & 99.50\% & 99.45\% & 99.80\% & 99.85\% & 99.60\%\\
		3 & 61.10\% & 93.25\% & 91.50\% & 4.55\% & 98.60\% & 97.55\%\\
		4 & 99.95\% & 99.95\% & 99.95\% & 99.95\% & 99.90\% & 99.85\%\\
		5 & 8.90\% & 55.65\% & 81.05\% & 2.55\% & 78.90\% & 98.45\%\\
		6 & 100.00\% & 100.00\% & 100.00\% & 100.00\% & 99.30\% & 98.60\% \\
		7 & 100.00\% & 100.00\% & 100.00\% & 100.00\% & 99.95\% & 99.90\%\\
		8 & 99.55\% & 99.50\% & 99.50\% & 99.65\% & 99.60\% & 99.45\% \\
		9 & 84.55\% & 94.70\% & 94.25\% & 8.35\% & 98.70\% & 97.55\% \\
		10 & 98.75\% & 98.75\% & 98.70\% & 96.40\% & 99.10\% & 98.65\% \\
		11 & 99.90\% & 99.90\% & 99.85\% & 99.45\% & 99.85\% & 99.80\%\\
		12 & 99.30\% & 99.10\% & 98.85\% & 60.50\% & 99.65\% & 99.20\%\\
		13 & 97.30\% & 97.20\% & 97.15\% & 97.60\% & 98.50\% & 97.40\% \\
		14 & 99.85\% & 99.80\% & 99.80\% & 99.90\% & 99.85\% & 99.80\%\\
		15 & 8.10\% & 61.60\% & 80.80\% & 1.55\% & 80.05\% & 96.05\% \\
		16 & 0.20\% & 72.20\% & 85.55\% & 0.85\% & 65.85\% & 97.30\%\\
		17 & 99.05\% & 99.00\% & 99.00\% & 99.15\% & 99.05\% & 98.95\%\\
		18 & 97.85\% & 97.80\% & 97.75\% & 92.10\% & 98.60\% & 97.75\%\\
		19 & 99.80\% & 99.75\% & 99.75\% & 99.90\% & 99.85\% & 99.75\%\\
		20 & 99.40\% & 99.35\% & 99.70\% & 99.40\% & 99.35\% & 99.10\%\\
		21 & 7.10\% & 72.60\% & 81.25\% & 2.15\% & 79.55\% & 97.35\%\\
		Average & 79.05\% & 92.35\% & 95.38\% & 69.70\% & 94.95\% & 98.66\%\\
		
		\bottomrule[1.5pt]
	\end{tabular}
	\label{FENet}
\end{table*}

\subsection{Extension of AE-FENet}
In addition, it should be noted that AE-FENet could be extensible. Regarding the feature transformation layer, alternative approaches such as Sparse Autoencoder (Sparse AE) \cite{Shao2017}, Autoencoder with Attention (Attention AE) \cite{LiuXing2018}, and Variational Autoencoder (VAE) \cite{vahdat2020nvae} can be employed as substitutes for the autoencoder.

\subsubsection{Sparse Autoencoder}
Sparse AE incorporates sparse constraints into the loss function, building upon the previous work \cite{srivastava2014dropout}, \cite{liu2016multimodal}. This variant of the autoencoder utilizes a neural network with a substantial number of neurons. Even when a significant portion of neurons in the hidden layer are inhibited, the sparse AE effectively captures the input data's features. To enforce sparsity, the activation value of the hidden layer is penalized using KL divergence, encouraging the hidden layer's code to converge towards zero.

Here the sparse parameter $\rho$ means the expected average neuron activation value. The average activation degree $\rho_{i}$ of the $i$-th neuron node can be calculated by:
\begin{equation}
\hat{\rho_{j}} = \frac{1}{m} \sum_{i=1}^{m}[a_{j}^{(2)} \bm{x}_{i}]
\end{equation}	
where $m$ is the number of input variables, $a_{j}^{(2)} \bm{x}_{i}$ represents the activation value of the $j$-th neuron under the input of $\bm{x}_{i}$. Therefore, the KL divergence can be expressed by:
\begin{equation}
KL(\rho||\hat{\rho_{j}}) = \rho log\frac{\rho}{\hat{\rho_{j}}} + (1-\rho)log\frac{1-\rho}{1-\hat{\rho_{j}}}.
\end{equation}	
Using the sparse option to regularize, the loss function is expressed as follows:
\begin{equation}
\bm{J}_{sparse}(W,b) = \bm{J}(W,b)  + \beta \sum_{j=1}^{s}KL(\rho ||\hat{\rho_{j}})
\end{equation}	
where $\beta$ is the weight of KL divergence. 

The AE-FENet associated with the sparse autoencoder implementation is referred to as sparse AE-FENet. Table \ref{otherAE} contains the FDRs of the sparse AE-FENet for common faults in TEP. Notably, the average FDR achieved an impressive level of 94.57\%, underscoring the efficacy of the sparse AE-FENet.

\subsubsection{Autoencoder with Attention}
The mechanism of attention draws inspiration from the human visual system and finds extensive applications in language text and image processing \cite{Shi2020}. Within the attention model, the input is no longer encoded as a static intermediate vector but is instead treated as a sequence of vectors by the encoder. Self-attention is derived from the hidden layer code of a time series, specifically the batch size, and is incorporated into the original code. The incremental selection and addition of attention to the sequence result in enhanced network focus and improved utilization.

The corresponding AE-FENet is referred to as attention AE-FENet when the attention mechanism is employed. The FDRs (False Discovery Rates) of attention AE-FENet are listed in Table \ref{otherAE} for a selected batch size of 10. Notably, attention AE-FENet demonstrates exceptional performance in detecting typical faults in TEP. In this context, three feature transformation layers are utilized to enhance the overall detection capabilities.

\subsubsection{Variational Autoencoder}
VAE refers to the idea of parameter estimation, assuming that the hidden layer encoding of the training data is not a fixed value but a random variable, which obeys a certain distribution. The input of the decoder is sampled from $z$, which is defined by the mean vector $\mu$ and the standard deviation vector $\sigma$. The random vector $\xi$ obeys normal distribution $N(0,1)$, $z = \mu + \sigma \cdot \xi$. By learning from a random distribution, the training process incorporates a higher degree of randomness, effectively mitigating overfitting to the training data.

The loss function of VAE comprises two primary components: a reconstruction loss and the KL divergence. The reconstruction loss serves the same purpose as in an autoencoder, capturing the discrepancy between the input and the output of the VAE. The KL divergence term is employed to enforce the latent space z to adhere to a normal distribution. Thus, the chosen loss function can be expressed as follows:
\begin{equation}
\mathcal{L}(\theta,\phi;x) = E_{Q_{\phi}}[logP_{\theta}(x|z)] - D_{KL}(Q_{\phi}(z|x)||P_{\theta}(z))
\end{equation}	

The correponding AE-FENet variant in this context is referred to as attention VAE-FENet. The FDRs of VAE-FENet for typical faults in TEP is showed in Table \ref{otherAE}. Here VAE-FENet$^{1}$ uses seven basic detectors mentioned before and VAE to construct basic detectors, whlie VAE-FENet$^{2}$ only include VAE as basic detectors. In Table \ref{otherAE}, when $l^{max}=2$, the FDRs of VAE-FENet are all larger than 95\%. For VAE-FENet$^{1}$, the FDR of fault 9 is 98.45\%, and that of fault 15 is 96.30\%. These results further demonstrate the scalability of the proposed AE-FENet when different types of autoencoders are utilized in the hidden feature transformation layers.

\begin{table*}
	\centering
	\caption{FDRs(\%) of sparse AE-FENet, attention AE-FENet, VAE-FENet$^{1}$ and VAE-FENet$^{2}$}
	\begin{tabular}{ccccccccccccc}
		\toprule[1.5pt]
		\multirow{2}*{Fault} & \multicolumn{3}{c}{Sparse AE-FENet} & \multicolumn{3}{c}{Attention AE-FENet} & \multicolumn{3}{c}{VAE-FENet$^{1}$} & \multicolumn{3}{c}{VAE-FENet$^{2}$}\\
		\cmidrule(r){2-4} \cmidrule(r){5-7} 
		\cmidrule(r){8-10} \cmidrule(r){11-13}
		~ & $l^{max}$=0 & $l^{max}$=1 & $l^{max}$=2 & $l^{max}$=1 & $l^{max}$=2 & $l^{max}$=3 & $l^{max}$=0 & $l^{max}$=1 & $l^{max}$=2 & $l^{max}$=0 & $l^{max}$=1 & $l^{max}$=2 \\
		\midrule
		3 & 4.55\% & 98.55\% & 97.70\% & 96.35\% & 97.10\% & 95.90\% & 4.55\% & 97.45\% & 98.25\% & 0.90\% & 55.60\% & 99.50\% \\ 
		
		5 & 2.55\% & 62.40\% & 88.65\% & 39.50\% & 85.70\% & 90.30\% & 2.55\% & 64.65\% & 98.65\% & 1.70\% & 61.10\% & 99.10\% \\ 
		
		9 & 8.35\% & 98.70\% & 98.25\% & 98.30\% & 98.25\% & 97.25\% & 8.35\% & 99.10\% & 98.45\% & 1.25\% & 27.25\% & 96.40\%\\ 
		
		15 & 1.55\% & 62.50\% & 96.25\% & 36.65\% & 87.05\% & 92.10\% & 1.55\% & 69.55\% & 96.30\% & 0.50\% & 35.75\% & 98.45\%\\ 
		
		16 & 0.85\% & 50.30\% & 92.95\% & 18.50\% & 76.65\% & 86.20\% & 0.85\% & 49.45\% & 96.95\% & 0.70\% & 30.85\% & 96.30\%\\ 
		
		21 & 2.15\% & 63.70\% & 93.65\% & 32.50\% & 86.10\% & 87.50\% & 2.15\% & 64.05\% & 96.95\% & 1.45\% & 45.30\% & 99.05\% \\ 
		
		Average & 3.33\% & 72.68\% & 94.57\% & 53.63\% & 88.48\% & 91.54\% & 3.33\% & 74.04\% & 97.59\% & 1.08\% &42.64\% & 98.13\%\\	
		\bottomrule[1.5pt]
	\end{tabular}
	\label{otherAE}
\end{table*}

\section{Conclusion}

The paper introduces a novel approach called AE-FENet as an enhancement to the original FENet. AE-FENet utilizes an autoencoder instead of a simple PCA in the hidden feature transformation layers of FENet. Consequently, AE-FENet belongs to the category of unsupervised deep learning networks and effectively extracts deep features related to incipient faults, including faults 3, 9, and 15. In addition to its superior detection performance, AE-FENet exhibits high scalability by incorporating various types of autoencoders such as sparse, attention, and variational autoencoders. This flexibility enables its application in diverse industrial processes, making it suitable for comprehensive monitoring purposes. 

\bibliography{refs} 

\end{document}